\documentclass[11pt,a4paper]{article}

\usepackage{mathpple}
\usepackage{graphicx}
\usepackage{url}
\usepackage{paralist}

\usepackage{geometry}
\begin{document}

\noindent
\textbf{Preprint of:}\\
G. Kn\"{o}ner, S. Parkin, T. A. Nieminen, N. R. Heckenberg
and H. Rubinsztein-Dunlop\\
``Forces from highly focused laser beams:
modeling, measurement and application to refractive index measurements''\\
in R. Sang and J. Dobson (eds),
\textit{Australian Institute of Physics (AIP)
  17th National Congress 2006: Refereed Papers},
Australian Institute of Physics, 2006 (CD-ROM, unpaginated).

\hrulefill

\begin{center}

{\huge
\textbf{Forces from highly focused laser beams:
modeling, measurement and application to refractive index measurements}}

\vspace{2mm}

\textit{G. Kn\"{o}ner, S. Parkin, T. A. Nieminen, N. R. Heckenberg
and H. Rubinsztein-Dunlop}\\
School of Physical Sciences, The University of Queensland, Australia

\subsection*{Abstract}

\end{center}

\begin{quote}
The optical forces in optical tweezers can be
robustly modeled over a broad range of parameters
using generalsed Lorenz--Mie theory. We describe the procedure,
and show how the combination of experimental measurement
of properties of the trap coupled with computational modeling,
can allow unknown parameters of the particle---in this case, the
refractive index---to be determined.
\end{quote}

\section*{Introduction}

Light carries momentum, and changes in momentum equal applied forces.
Focusing laser beams to small spot sizes creates high intensities and
makes the momentum carried by the light comparable to other the other
forces acting at that scale. The momentum can thus be exploited for
applications ranging from atom trapping for Bose--Einstein condensation
to molecule and nano-particle trapping to the trapping of entire live
cells. The interaction of light with small and large particles is
relatively easy to describe with Rayleigh
scattering and geometrical optics, respectively. For the important
particle size range with radii from 0.1 to 5 times the laser wavelength,
direct solution of either the Maxwell equations or the vector
Helmholtz equation is required. For the case of a spherical particle,
an analytical solution is available: Lorenz--Mie theory
(Lorenz~1890; Mie~1908).

We show how forces on particles in laser traps can be robustly modeled
for a wide range of parameters by employing generalised Lorenz--Mie
theory. We present results in the form of parameter landscapes which are of
interest for a broader audience.

We compare computational modeling with experimental measurement of
the forces acting in an optical trap, finding excellent agree between
precision measurements of the
optical spring constant and the theoretical predictions. We use the
combination of such measurements and theoretical modeling to determine
the refractive index of a microparticle.

\section*{Computational Modeling of Optical Tweezers}

A general divergence-free solution of the vector Helmholtz
equation can be written in terms of vector spherical wavefunctions:
\begin{eqnarray}
\mathbf{M}_{nm}^{(1,2)}(k\mathbf{r}) & = & N_n h_n^{(1,2)}(kr)
\mathbf{C}_{nm}(\theta,\phi) \\
\mathbf{N}_{nm}^{(1,2)}(k\mathbf{r}) & = & \frac{h_n^{(1,2)}(kr)}{krN_n}
\mathbf{P}_{nm}(\theta,\phi) + N_n
\left( h_{n-1}^{(1,2)}(kr) -
\frac{n h_n^{(1,2)}(kr)}{kr} \right) \mathbf{B}_{nm}(\theta,\phi)
\nonumber
\end{eqnarray}
where $h_n^{(1,2)}(kr)$ are spherical Hankel functions of the first and second
kind,
$N_n = [n(n+1)]^{-1/2}$ are normalization constants, and
$\mathbf{B}_{nm}(\theta,\phi) = \mathbf{r} \nabla Y_n^m(\theta,\phi)$,
$\mathbf{C}_{nm}(\theta,\phi) = \nabla \times \left( \mathbf{r}
Y_n^m(\theta,\phi) \right)$, and
$\mathbf{P}_{nm}(\theta,\phi) = \hat{\mathbf{r}} Y_n^m(\theta,\phi)$
are the vector spherical
harmonics (Mishchenko~1991),
and $Y_n^m(\theta,\phi)$ are normalized scalar spherical harmonics. The usual
polar spherical coordinates are used, where $\theta$ is the co-latitude
measured
from the $+z$ axis, and $\phi$ is the azimuth, measured from the $+x$ axis
towards the $+y$ axis.

In general, there will be an incoming part of the field:
\begin{equation}
\mathbf{E}_\mathrm{in} = \sum_{n=1}^\infty \sum_{m = -n}^n
a_{nm} \mathbf{M}_{nm}^{(2)}(k\mathbf{r}) +
b_{nm} \mathbf{N}_{nm}^{(2)}(k\mathbf{r}),
\label{incoming_expansion}
\end{equation}
and an outgoing part:
\begin{equation}
\mathbf{E}_\mathrm{out} = \sum_{n=1}^\infty \sum_{m = -n}^n
p_{nm} \mathbf{M}_{nm}^{(1)}(k\mathbf{r}) +
q_{nm} \mathbf{N}_{nm}^{(1)}(k\mathbf{r}).
\label{outgoing_expansion}
\end{equation}
The fields can be compactly described by arranged the mode
amplitude coefficients $a_{nm}$ and $b_{nm}$ as components
of an incoming amplitude vector $\mathbf{a} =
[ a_{0,-1}, a_{0,0}, a_{0,+1}, .. , b_{0,-1}, b_{0,0}, b_{0,+1}, .. ]$,
and $p_{nm}$ and $q_{nm}$ as an outgoing amplitude vector $\mathbf{p}$.
If the electromagnetic properties of the scatterer are linear, these
two will be related by a linear transformation
\begin{equation}
\mathbf{p} = \mathbf{T} \mathbf{a}.
\end{equation}
Here, the matrix $\mathbf{T}$ is called the \emph{transition matrix}
or \textit{T}-matrix. In principle, the field expansions and the
\textit{T}-matrix are infinite, but, in practice, can safely be
truncated at a finite $n_\mathrm{max}$, typically with
$n_\mathrm{max} \approx kr_0$, where $r_0$ is a radius that
enclosed the particle or the beam waist.

When the particle is a homogeneous isotropic
sphere, the \textit{T}-matrix is diagonal, with elements given
by the analytical Lorenz--Mie solution (Lorenz~1890; Mie~1908;
van de Hulst~1981). For non-spherical particles, the \textit{T}-matrix
can still be calculated, but is a more computationally intensive
task (Nieminen~\textit{et~al.}~2003b).

The \textit{T}-matrix need only be calculated once for each particle.
It is a complete description of the scattering properties of the
particle at that wavelength, with all information about the incident field
contained in $\mathbf{a}$. If $\mathbf{a}$ and the \textit{T}-matrix
are known, then $\mathbf{p}$ can be found. At this point, the fields
outside the particle are known, and can be used to find the momentum
and angular momentum fluxes of the incoming and outgoing fields, with
the optical force and torque being given by the differences between them.
While one might guess that this would require numerical integration
of the Poynting vector over a surface enclosing the particle, the
orthogonality properties of the spherical functions involved can be used to
reduce this to a sum of products of the mode amplitudes
(Farsund and Felderhof~1996; Crichton and Marston~2000).

Finally, we need to consider how the incident field mode amplitudes
can be found. This is a far from simple task; our method is to
use an overdetermined point-matching algorithm in the far field
(Nieminen~\textit{et~al.}~2003a). This is most simply done
in a coordinate system with the origin at the focus and the beam
axis coincident with the $z$-axis, in which case the rotation symmetry
of the beam can be used to greatly reduce the computational requirements.
An example of the instantaneous fields of a beam calculated in this
manner is shown in figure 1.

\centerline{\includegraphics[width=0.48\columnwidth]{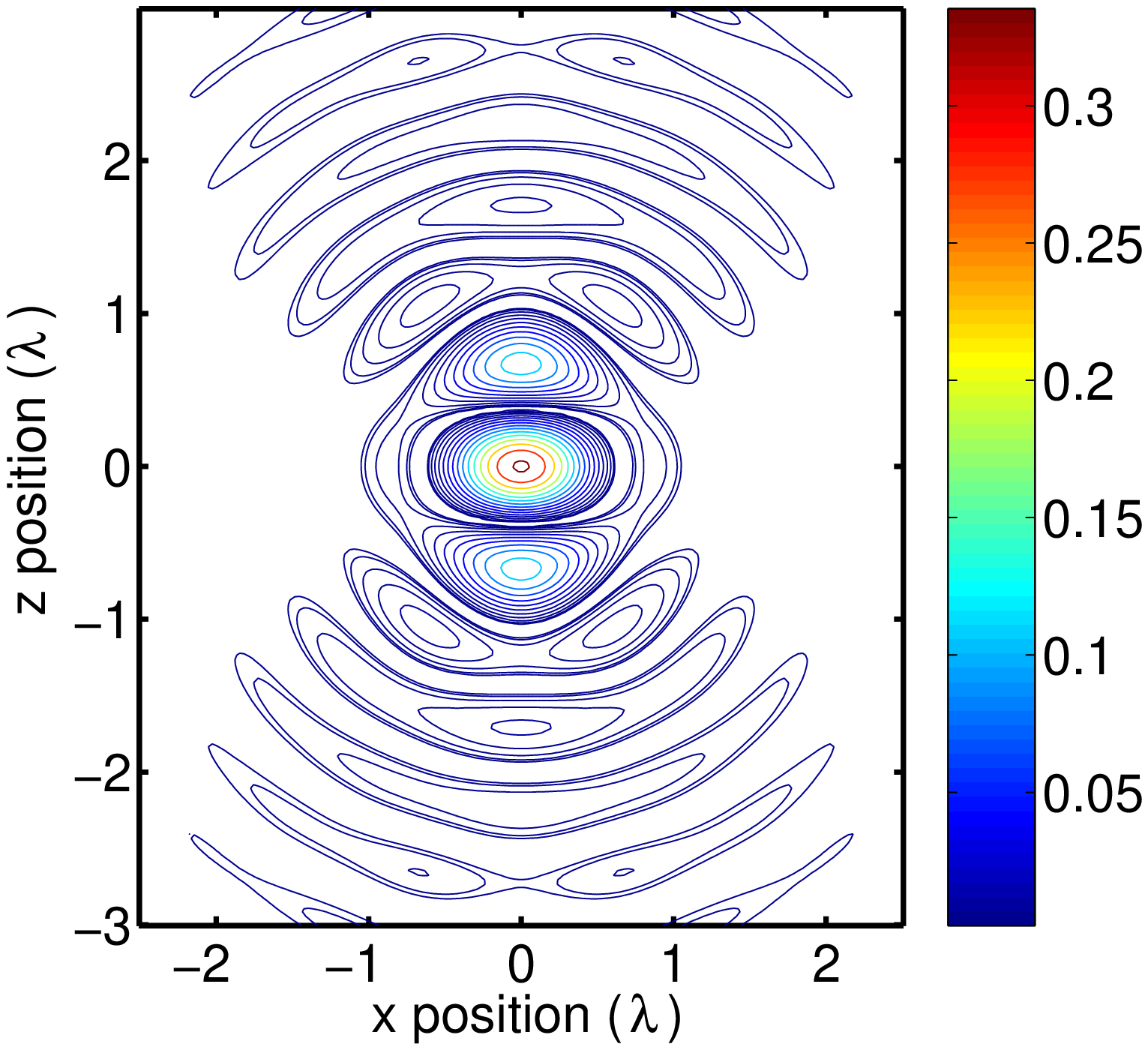}}
~\vspace{-16mm}~\\

\textbf{Figure 1. Instantaneous fields of a tightly focused beam}

The mode amplitude coefficients of the beam in other coordinate
systems can be found using the translation and rotation addition
theorems for vector spherical wavefunctions
(Choi~\textit{et~al.}~1999; Videen~2000; Gumerov and Duraiswami~2003).

\includegraphics[width=0.45\columnwidth]{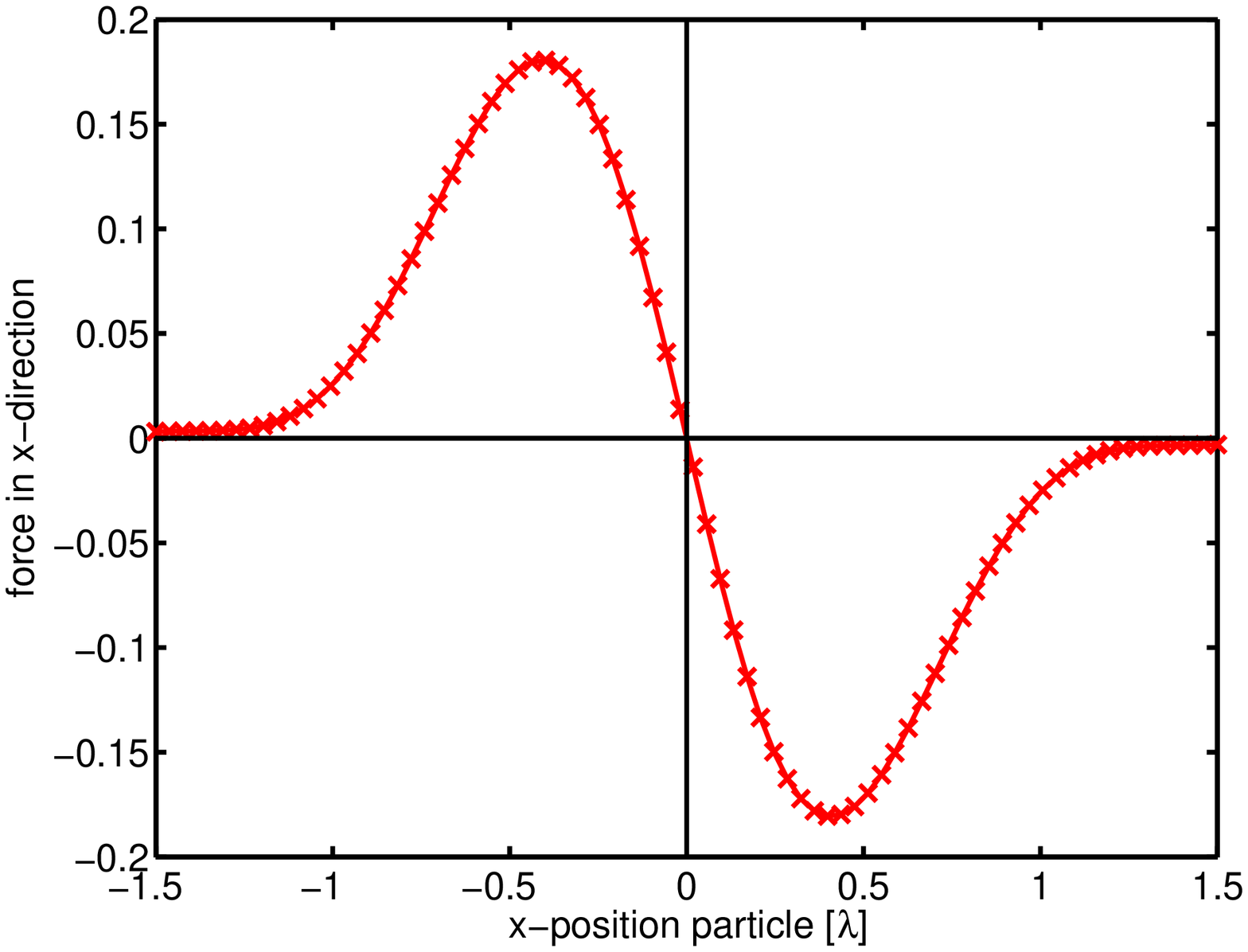}
\hfill
\includegraphics[width=0.45\columnwidth]{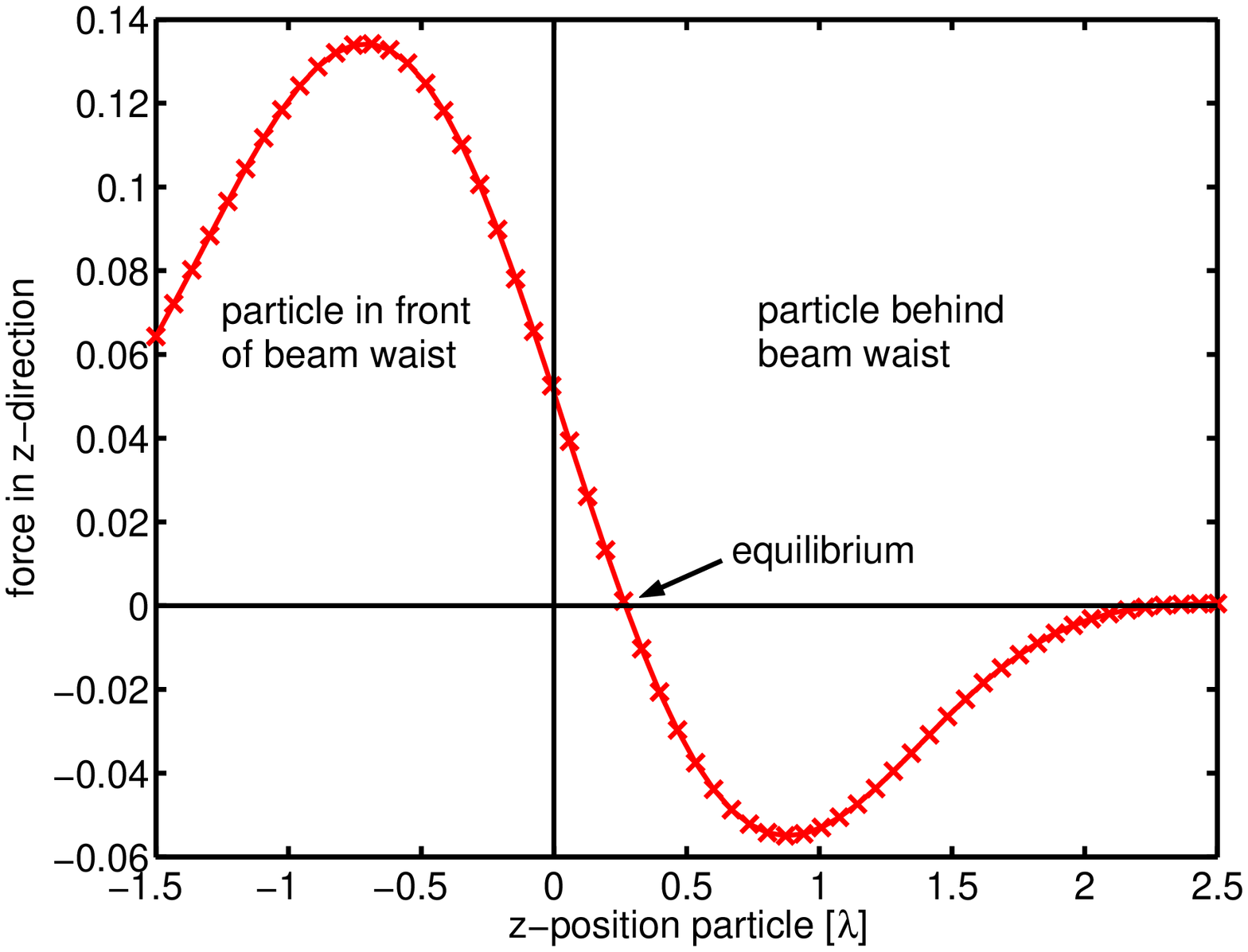}

\textbf{Figure 2. Typical force curves}

The typical behaviour of the optical force as a function of radial and
axial position within the trap is shown in figure~2. The radial force
is symmetric about the beam axis, while the axial force is asymmetric
about the focal plane; the gradient force acts towards the focal plane
for displacements in either direction, while forces due to reflection
of the trapping beam from the particle always act in the direction
of propagation. Thus, the equilibrium position of the particle is somewhat
past the focus. If the reflection force (ie, the force usually
called the ``scattering force'', although it should be recognised that
both this force \emph{and} the gradient force arise through scattering)
exceeds the maximum gradient force, trapping will not be possible.
This will occur for high refractive index particles. The maximum axial
restoring force can be calculated, and the parameters for which
particles can be trapped can be determined. Figure 3 shows the combinations
of size and refractive index for which particles can be trapped.

\centerline{\includegraphics[angle=270,width=0.65\textwidth]{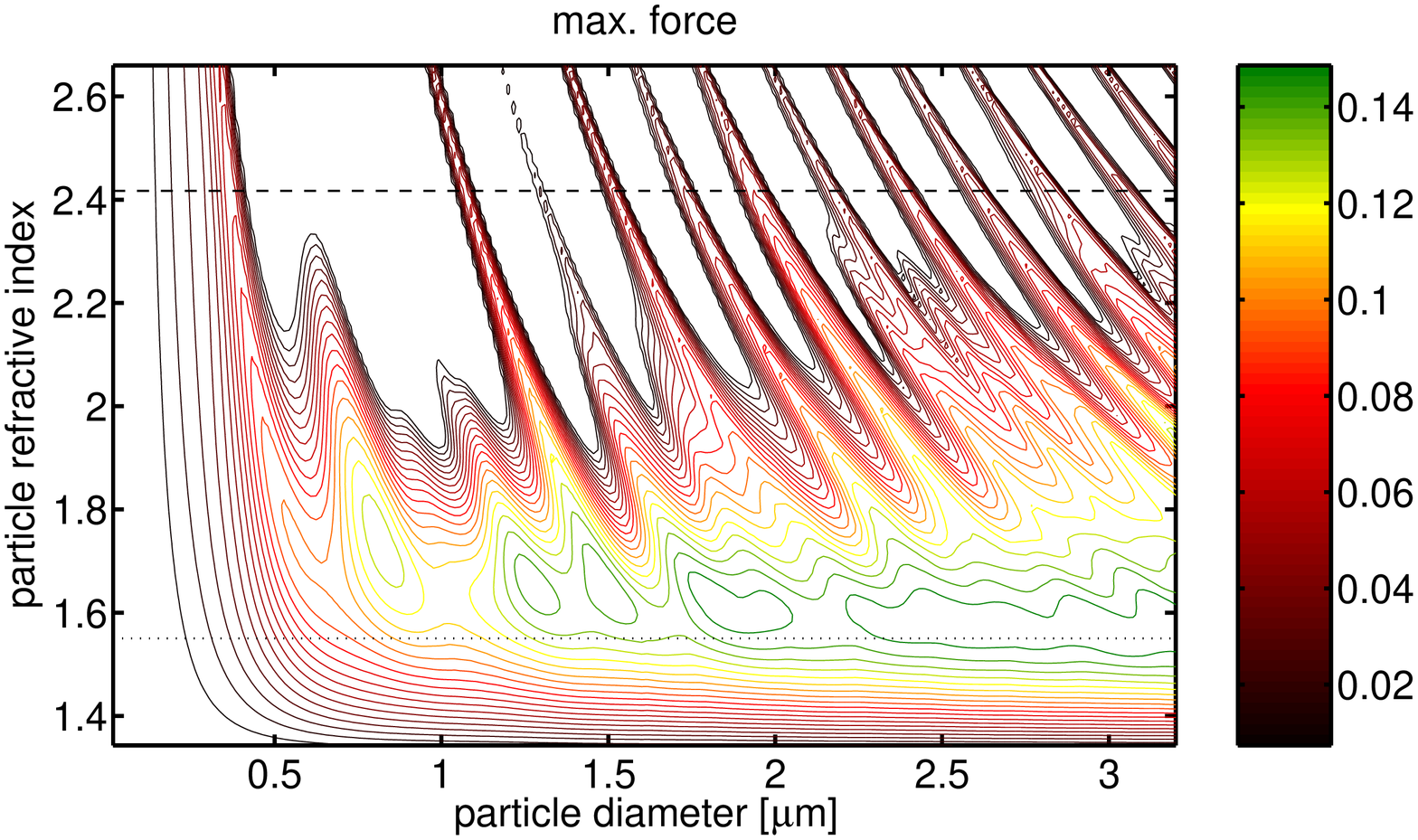}}

\textbf{FIgure 3. Maximum axial restoring force.} Where contours
are absent, the reflection force overcomes the gradient force,
and trapping is not possible.

\section*{Comparison with Experiment and Refractive Index Measurement}

Precision measurement of the properties of an optical trap allows the
above modeling methodology to be tested. The spring constant of the
trap was measured for a range of microspheres (silica, PMMA, and
polystyrene). The silica microspheres were used to determine the
laser power at the focus of the trap, and this power was then used to
calculate the spring constants for the other microspheres as a
function of refractive index. This is shown in figure 4. Excellent
agreement was obtained between the refractive indices as indicated
by comparison of the measured spring constants and the theoretical
curves, and the known refractive indices .

\includegraphics[width=0.45\columnwidth]{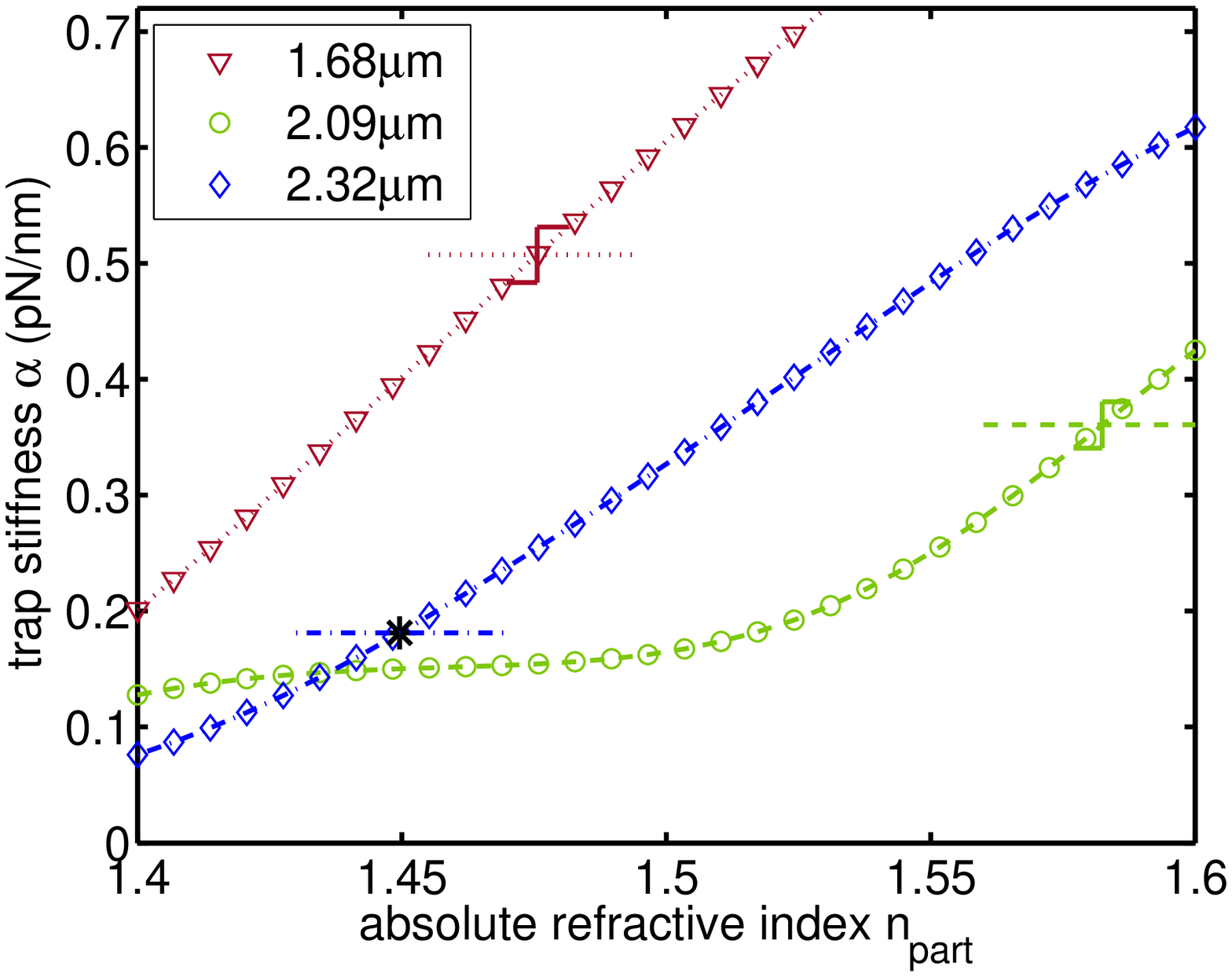}
\hfill
\includegraphics[width=0.45\columnwidth]{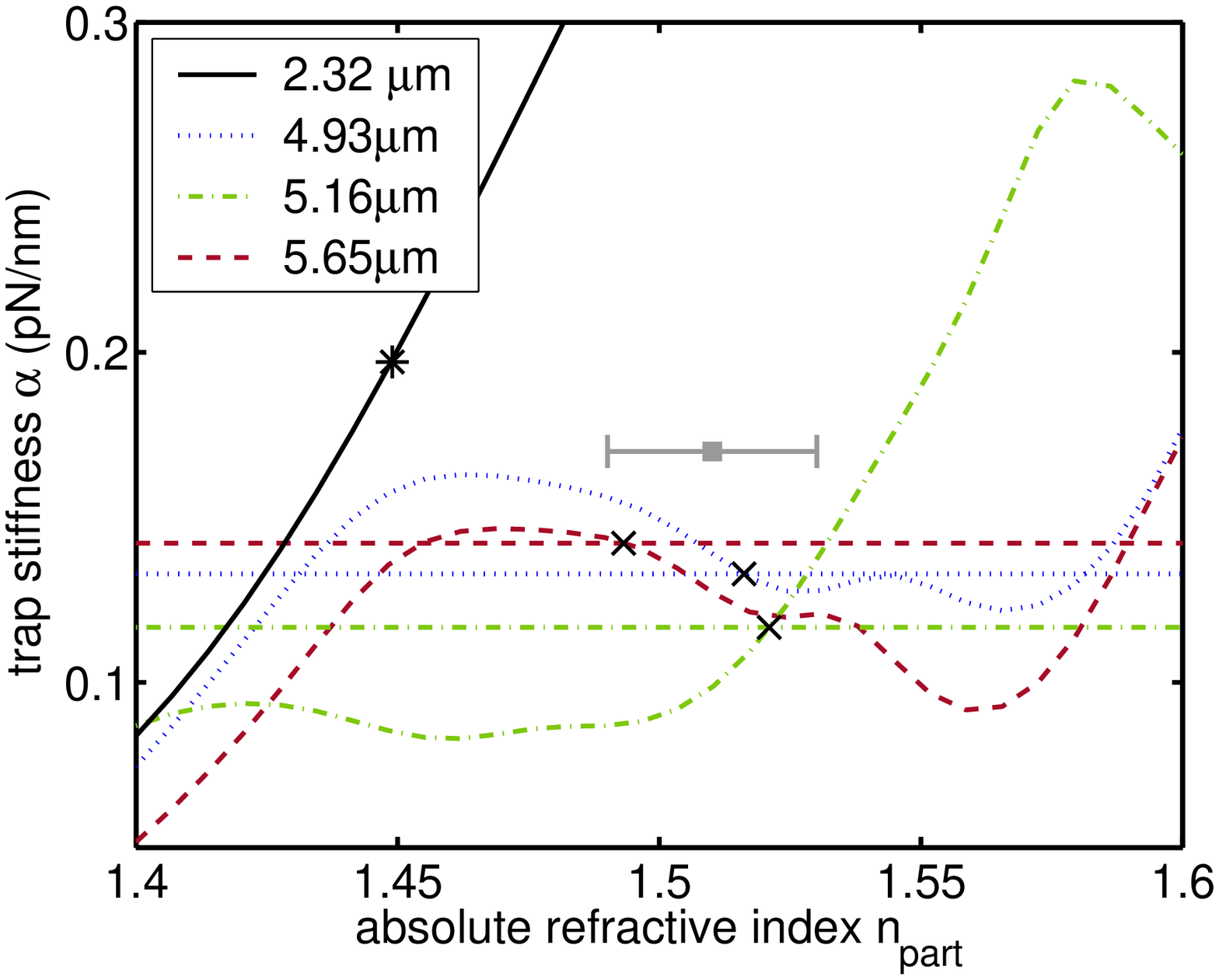}\\
~\vspace{-14mm}~\\
(a) \hfill (b) \hfill ~

\textbf{Figure 4. Spring constants---measurement and calculation.}
(a) shows calculated spring constants and experimental
measurements for three different types of microspheres:
silica (red), PMMA (blue), and polystyrene (green). (b) shows
a calibration curve for a known microsphere (silica; black) and
calculated and measured spring constants for organosilica
microspheres of unknown refractive index.

The refractive index of micrometre sized objects is an important
quantity, strongly affecting their optical properties.
However, it is not easily measured, especially for
particles for which there is no equivalent bulk material, or which
must remain in a particular environment to avoid alteration of
optical properties, ruling out the possibility of index matching.
Most methods based on scattering require a monodisperse sample---since
the method presented here uses only a single particle at a time,
a polydisperse sample presents no undue difficulty, and can even be
an advantage, as some sizes in the range present may allow more
accurate determination of the refractive index. Also, as a
single-particle method, there is no need to account for complications
such as multiple scattering.
Our method of testing the accuracy of the modeling can be directly
applied to the measurement of the refractive index of
particles for which it is unknown. Accordingly, 
we measured the spring constants for different sizes of
organosilica particles, and determined the refractive
index to be $1.51\pm 0.02$ (Kn\"{o}ner~2006).

\section*{Refererences}

Choi, C.H., Ivanic, J., Gordon, M.S. and Ruedenberg, K. (1991).
Rapid and stable determination of rotation matrices between
spherical harmonics by direct recursion.
{\em Journal of Chemical Physics} {\bf 111}, 8825-31.\\
Crichton, J.H. and Marston, P.L. (2000).
The measurable distinction between the spin and orbital angular
momenta of electromagnetic radiation.
{\em Electronic Journal of Differential Equations} {\bf Conf. 04}, 37-50.\\
Farsund, {\O}. and Felderhof, B.U.
Force, torque, and absorbed energy for a body of arbitrary shape and
constitution in an electromagnetic radiation field.
{\em Physica A} {\bf 227}, 108-30.\\
Gumerov, N.A. and Duraiswami, R. (2003).
Recursions for the computation of multipole translation and
rotation coefficients for the 3-{D} {H}elmholtz equation.
{\em SIAM Journal on Scientific Computing} {\bf 25}, 1344-81.\\
Kn\"{o}ner, G., Parkin, S., Nieminen, T.A., Heckenberg, N.R. and
Rubinsztein-Dunlop, H. (2006).
Measurement of refractive index of single microparticles.
\textit{Physical Review Letters} \textbf{97}, 157402.\\
Lorenz, L. (1890).
Lysbev{\ae}gelsen i og uden for en af plane Lysb{\o}lger belyst Kugle.
\textit{Videnskabernes Selskabs Skrifter} \textbf{6}, 2-62.\\
Mie, G. (1908).
Beitr\"{a}ge   zur   Optik   tr\"{u}ber   Medien,   speziell  kolloidaler
Metall\"{o}sungen. \textit{Annalen der Physik} \textbf{25}, 377-445.\\
Mishchenko, M.I. (1991).
Light scattering by randomly oriented axially symmetric particles.
{\em Journal of the Optical Society of America A} {\bf 8}, 871-82.\\
Nieminen, T.A., Rubinsztein Dunlop, H. and Heckenberg, N.R. (2003a).
Multipole expansion of strongly focussed laser beams.
\textit{Journal of Quantitative Spectroscopy and Radiative Transfer}
\textbf{79-80}, 1005-17.\\
Nieminen, T.A., Rubinsztein Dunlop, H. and Heckenberg, N.R. (2003b).
Calculation   of   the  \textit{T}-matrix:   general  considerations  and
application of the point-matching method.
\textit{Journal of Quantitative Spectroscopy and Radiative Transfer}
\textbf{79-80}, 1019-29.\\
van de Hulst, H.C. (1981), Light scattering by small particles.
Dover, New York.\\
Videen, G. (2000).
Light scattering from a sphere near a plane interface. In
{\em Light Scattering from Microstructures},  Moreno, F. and Gonz\'{a}lez,
F., eds., {\em Lecture Notes in Physics} \textbf{534}, p.81-96,
Springer-Verlag, Berlin.

\end{document}